\documentclass{elsart}

 \usepackage{graphicx}

\usepackage{amssymb}

\begin{document}

\begin{frontmatter}



\title{Bardeen-Cooper-Schrieffer Formalism of Superconductivity in Carbon Nanotubes}


\author{Shi-Dong Liang}

\address{State Key Laboratory of Optoelectronic Material and Technology and
Guangdong Province Key Laboratory of Display Material and Technology,
School of Physics and Engineering, 
Sun Yat-Sen (Zhongshan) University, Guangzhou, 510275, People's Republic of China
}

\begin{abstract}
We develop the Bardeen-Cooper-Schrieffer (BCS) formalism for the superconductivity of carbon nanotubes. It is found that 
the superconducting transition temperature $T_{c}$ of single-wall carbon nanotubes decreases exponentially 
with the increase of the tube diameter because the density of states near the Fermi energy is inverse
proportional to the tube diameter. For the multi-wall carbon nanotubes, 
the Cooper paring hopping between layers enhances the superconducting 
correlation and increase the superconducting transition temperature, 
which is consistent with the experimental observation. 
\end{abstract}

\begin{keyword}
BCS theory; Carbon nanotube
\PACS 74.20.Fg;	73.63.Fg
\end{keyword}
\end{frontmatter}

\section{Introduction}
\label{}
Carbon nanotubes (CNs) have promised a great potential application in nanotechnology.\cite{Iijima} 
In recent years, superconductivity has been observed experimentally in the single-wall carbon nanotubes (SWCN),
the rope of SWCNs,\cite{Kasumov} and the multi-wall carbon nanotubes (MWCN)\cite{Takesue} by the low-temperature transport measurements. 
For the individual SWCN and the rope of SWCN, the superconducting transitions 
have been observed at a temperature below $1$K.\cite{Kasumov} For the end-bonded MWCNs,
the superconducting transition temperature reaches $6\sim12K$,\cite{Takesue}
which is much higher than that of the individual SWCN and the rope of SWCN.
This implies that the interlayer coupling may enhance the superconducting correlation.\cite{Takesue} 
The investigation of the magnetic properties reveals that the superconductivity emerges at temperatures
below $15K$ for a small diameter tube (4 angstrom) embadded in a zeolite.\cite{Tang}

From a theoretical point of view, it is still not very clear how to understand the mechanism of superconductivity
in CNs.  Generally, CNs may be regarded as a one-dimensional material (1D). The 1D systems face some obstructions
that prevent the emergence of superconductivity, such as Luttinger liquid states with a repulsive electron-electron
interaction\cite{Egger} and Peierls phase transition. However, one believes     
that the proximity effect could play an essential role in superconductivity state in CNs.\cite{Kasumov} 
Another scenario on superconductivity of CNs is obtained by the effective low-energy theory for interacting 
electrons in metallic SWCN, which predicts that SWCNs can be described as a Luttinger 
liquid with an attractive electron-electron interaction.\cite{Martino} 
The breathing modes specific to CNs can be the origin of a strong electron-phonon coupling giving rise to
attractive electron-electron interactions.\cite{Martino} 
A short-range attractive electronic interaction originates from the exchange of phonons.\cite{Gonzalez}
The single-particle hopping between neighboring nanotubes in a rope is strongly suppressed
because the different helical structure of nanotubes, in general, leads to the mismatching of the atoms
between tubes. The intertube coherence is established mainly through the tunneling of Cooper pairs, 
which enhances the superconducting correlation.\cite{Gonzalez}
Exactly speaking, CN is a quasi-one-dimensional system. The energy band contains some split bands near Fermi
level. The interlayer coupling in MWCNs could play an important role in superconductivity of CNs.
Superconductivity in CNs should be intrinsic \cite{Martino,Gonzalez} and 
the nature of superconductivity in CNs is still attributed to Bardeen-Cooper-Schrieffer(BCS) mechanism.\cite{BCS}  
Thus, the BCS formalism for CNs is still an interesting issue.

In this paper, we will develop a formalism of MWCNs based on BCS theory, 
and discuss the superconducting correlation and the critical temperature. In the section II, we will propose
a BCS Hamiltonian of MWCNs in the tight-binding approximation. We will give a mean field solution
of this Hamiltonian, the self-consistent equations on the superconducting energy gap and the superconducting transition
temperature of MWCN in the section III. Finally, we will give a discussion and conclusion.

\section{BCS Hamiltonian}
Since the intrinsic superconductivity is observed in SWCNs one may believe
that the effective electron-electron interaction is attractive to form the Cooper pairs in the SWCN. 
We consider a general BCS Hamiltonian of MWCN constructed from a coupling multi-layer BCS Hamiltonian.\cite{George} 
The single electron hopping between layers is estimated to be much weaker than the Cooper pair hopping 
between layers due to the mismatching of atoms between layers.\cite{Gonzalez,Kasumov}
The Cooper pair hopping between layers dominates the interaction between
layers.\cite{Kasumov} Thus, the BCS Hamiltonian of MWCNs may be expressed in
terms of the coupling SWCN BCS Hamiltonian. 
\begin{equation}
H=\sum_{\ell}H_{\ell}(k)+\sum_{\langle \ell,\ell'\rangle}H_{\ell,\ell'},
\label{H}
\end{equation}
where
\begin{equation}
H_{\ell}(k)=\sum_{k,\sigma}E_{\ell}(k)c^{\dagger}_{k,\ell,\sigma}c_{k,\ell,\sigma}
-\sum_{k,k',\sigma}V_{\ell}c^{\dagger}_{k,\ell,\sigma}c^{\dagger}_{k,\ell,\bar{\sigma}}
c_{k',\ell,\bar{\sigma}}c_{k',\ell,\sigma},
\label{Hl}
\end{equation}
is the Hamiltonian of the $\ell$th tube. 
The $V_{\ell}$ in the second term of Eq.(\ref{Hl}) describes the intratube effective electron-electron interaction.
The $H_{\ell,\ell'}$ in the second term of  Eq.(\ref{H}) is the coupling between 
layers, where the sum $\langle\ell,\ell'\rangle$ runs only for the nearest layers in the tight-binding approximation. 
The layer-layer coupling Hamiltonian may be written as
\begin{equation}
H_{\ell,\ell'}=-\sum_{k,k',\sigma}V_{\ell,\ell'}c^{\dagger}_{k,\ell,\sigma}c^{\dagger}_{k,\ell,\bar{\sigma}}
c_{k',\ell',\bar{\sigma}}c_{k',\ell',\sigma}
\label{Hll}
\end{equation}
where the  
$V_{\ell,\ell'}$ in Eq.(\ref{Hll}) represents the intertube electron-electron interaction 
between the $\ell$th and $\ell'$th layers, which induces the Cooper pairing hopping between layers
and are measured relative to the corresponding screened Coulomb potential. 
The $c_{k,\ell,\sigma}^{\dagger }(c_{k,\ell,\sigma})$ is the creation (annihilation) operator
at the layer $\ell$ with the spin $\sigma$, where the $\ell=1,2...M$ labels the layers. 
The $E_{\ell}(k)=\epsilon_{\ell}(k)-E_{F}$, where $E_{F}$ is the Fermi energy. We set $E_{F}=0$ for convenience in the following. 
The $\epsilon_{\ell}(k)$ is the energy dispersion relation of the layer $\ell$ for free electrons, 
which can be obtained by the tight-binding approximation\cite{Saito}
\begin{equation}
\epsilon_{\ell}({\bf k})=\pm t\sqrt{1+4\cos \left(\frac{\sqrt{3}k_{x}^{(\ell)}a}{2}\right)
\cos \left(\frac{k_{y}^{(\ell)}a}{2}\right)
+4\cos^{2}\left(\frac{k_{y}^{(\ell)}a}{2}\right)},
\label{epk}
\end{equation}
and their wave vectors are \cite{Liang}
\begin{equation}
\left( 
\begin{array}{c}
k_{x}^{(\ell)} \\
k_{y}^{(\ell)} 
\end{array}
\right) =\left( 
\begin{array}{c}
\frac{a}{2L_{\ell}}(m_{\ell}-n_{\ell})k_{\ell}+\frac{\sqrt{3}\pi a}{L_{\ell}^{2}}(n_{\ell}+m_{\ell})q_{s} \\
\frac{\sqrt{3}a}{2L_{\ell}}(m_{\ell}+n_{\ell})k_{\ell}+\frac{\pi a}{L_{\ell}^{2}}(n_{\ell}-m_{\ell})q_{s}
\end{array}
\right),
\label{kxy} 
\end{equation}
where $a=2.46\AA$ is the lattice constant of hexagons. The $t$ is the intralayer electronic hopping constant. 
The SWCN is identified by the 
chiral vector ${\bf C}_{h}=(n,m)$, which length $L_{\ell}=a\sqrt{n^2_{\ell}+m^2_{\ell}+n_{\ell}m_{\ell}}$. 
The $(k_{\ell},q_{\ell})$ in Eq.(\ref{kxy}) are quantum numbers of the $\ell$th tube.
The sum $\sum_k$ in Eqs.(\ref{Hl}) and (\ref{Hll}) runs all states of the nanotubes 
$\ell$ and $\ell'$ in the Briullien zone, which includes the $k_{\ell}$ values within 
$-\frac{\pi }{T_\ell}<k_{\ell}<\frac{\pi }{T_{\ell}}$ and  $q _{\ell}=0,1,...N_{\ell}-1$, 
where $T_{\ell}=\sqrt{3}L_{\ell}/d_{R_{\ell}}$ is the length of the translational vector 
of the $\ell$th tube being parallel to the tube axis
and $N_{\ell}=2L_{\ell}^{2}/a^{2}d_{R_{\ell}}$ is the number of hexagons per unit cell of the $\ell$th tube. 
The $d_{R_{\ell}}=gcd(2n+m,2m+n)$ means the greatest common divisor of $2n_{\ell}+m_{\ell}$ and $2m_{\ell}+n_{\ell}$.
For comensurate MWCNs the atoms between layers are matching such that all $T_{\ell}$ are equal.\cite{Liang2}
The energy band structure of SWNTs depends on the chirality of the tube. The SWCN may be
metallic for $n_{\ell}-m_{\ell}=3i$ or semiconducting for $n_{\ell}-m_{\ell}\neq3i$, where $i$ is an integer.\cite{Saito}

\section{Mean field solution}
Based on the BCS mean field idea, factorizing the potential terms in Eq.(\ref{Hl}) and (\ref{Hll}),
$c^{\dagger}_{k,\ell,\sigma}c^{\dagger}_{k,\ell,\bar{\sigma}}c_{k,\ell',\bar{\sigma}}c_{k,\ell',\sigma}
=\langle c^{\dagger}_{k,\ell,\sigma}c^{\dagger}_{k,\ell,\bar{\sigma}}\rangle
c_{k,\ell',\bar{\sigma}}c_{k,\ell',\sigma}+
c^{\dagger}_{k,\ell,\sigma}c^{\dagger}_{k,\ell,\bar{\sigma}}
\langle c_{k,\ell',\bar{\sigma}}c_{k,\ell',\sigma}\rangle-
\langle c^{\dagger}_{k,\ell,\sigma}c^{\dagger}_{k,\ell,\bar{\sigma}}\rangle
\langle c_{k,\ell',\bar{\sigma}}c_{k,\ell',\sigma}\rangle $,
we can obtain the mean field form of Hamiltonian
\begin{equation}
H=\sum_{\ell,k,\sigma}\epsilon_{\ell}(k)c^{\dagger}_{k,\ell,\sigma}c_{k,\ell,\sigma}
-\sum_{\ell,k,\sigma}\Delta_{\ell}(c^{\dagger}_{k,\ell,\sigma}c^{\dagger}_{k,\ell,\bar{\sigma}}+h.c.)
+\sum_{\ell,k}\Delta^{\dagger}_{\ell}S_{\ell k}
\label{Hd}
\end{equation}
where the pair correlation, $S_{\ell k}$, has been defined by 
\begin{equation}
S_{\ell k}=\langle c_{k,\ell,\sigma}c_{k,\ell,\bar{\sigma}}\rangle,
\end{equation}
where $\langle X \rangle$ is the grand canonical average of $X$. The superconducting energy gap may be expressed as \cite{George}
\begin{equation}
\Delta_{\ell}=\sum_{k}V_{\ell}S_{\ell k}+\sum_{\ell'\ne \ell}V_{\ell,\ell'}S_{\ell' k},  
\label{D1}
\end{equation}
Using the Bogoliubov transform,
\begin{equation}
\left( 
\begin{array}{c}
c_{k,\ell,\sigma}\\
c^{\dagger}_{-k,\ell,-\sigma}
\end{array}
\right) =\left( 
\begin{array}{cc}
u_{k} & v_{k} \\
-v_{k} & u_{k} 
\end{array}
\right)
\left( 
\begin{array}{c}
\gamma_{k,\ell} \\
\gamma^{\dagger}_{-k,\ell} 
\end{array}
\right)
\label{B} 
\end{equation}
to diagonalize the Hamiltonian in Eq.(\ref{Hd}), the Hamiltonian can be written as
\begin{equation}
H_{BCS}=E_{G}
+\sum_{\ell,k}\xi_{\ell}(k)(\gamma^{\dagger}_{k,\ell}\gamma_{k,\ell}+\gamma^{\dagger}_{-k,\ell}\gamma_{-k,\ell}),
\label{Hk}
\end{equation}
where the ground state energy is
\begin{equation}
E_{G}=\sum_{\ell,k}(\epsilon_{\ell}(k)-\xi_{\ell}(k)+\frac{\Delta^{2}_{\ell}}{2\xi_{\ell}(k)}),
\end{equation}
and the quasiparticle energy spectrun, 
\begin{equation}
\xi_{\ell}(k)=\sqrt{\epsilon^{2}_{\ell}(k)+\Delta_{\ell}^{2}}.
\label{Ek}
\end{equation}
where $\ell$ labels the tube, which may be regarded as different energy bands of quasiparticles. 
Making use of the Bogoliubov transform Eq.(\ref{B}) and considering the tight-binding approximation, 
the superconductng energy gap $\Delta_{\ell}$ satisfies a set of equations, which matrix form may be written as\cite{George}
\begin{equation}
\left( 
\begin{array}{ccccc}
\alpha_{1,1} & \alpha_{1,2} &0 &\cdots &  0  \\
\alpha_{2,1} & \alpha_{2,2} &\alpha_{2,3}  &\cdots &  \vdots \\
0 & \alpha_{3,2} & \ddots& \vdots & 0 \\
\vdots & \vdots  & \cdots & \alpha_{M-1,M-1} & \alpha_{M,M-1}  \\
0 & 0  & \cdots & \alpha_{M,M-1}   & \alpha_{M,M} \\
\end{array}
\right)
\left(
\begin{array}{ccccc}
\Delta_{1} \\
\Delta_{2} \\
\vdots \\
\Delta_{M-1} \\
\Delta_{M} \\
\end{array}
\right)=0
\label{Del}
\end{equation}
where
\begin{eqnarray}
\alpha_{\ell,\ell'}=\left\{
\begin{array}{lll}
1-V_{\ell}F(\Delta_{\ell}) & {\rm for\   } \ell'=\ell \\
-V_{\ell,\ell'}F(\Delta_{\ell'}) & {\rm for\   } \ell'=\ell\pm 1 \\
0   & {\rm others }
\end{array}
\right.
\label{alpha} 
\end{eqnarray}
with
\begin{equation}
F(\Delta_{\ell})=\sum_{q=0}^{N_{\ell}-1}\int_{-\frac{\pi}{T_{\ell}}}^{\frac{\pi}{T_{\ell}}} 
\frac{dk}{\xi_{\ell}(k)}\tanh\left( \frac{\xi_{\ell}(k)}{2k_{B}T}\right)
\label{F}
\end{equation}
At zero temperature, the Eq.(\ref{F}) may reduce to 
$F(\Delta_{\ell})=\sum_{q=0}^{N_{\ell}-1}\int_{-\frac{\pi}{T_{\ell}}}^{\frac{\pi}{T_{\ell}}} \frac{dk}{\xi_{\ell}(k)}$.

In principle, numerically solving Eqs.(\ref{Del}) and (\ref{F}) associated with Eqs.(\ref{epk}),(\ref{kxy}) and (\ref{Ek}), 
we can obtain the superconducting paring, superconducting transition temperature and their relationships. 
The $F(\Delta_{\ell})$ should satisfy Eqs.(\ref{Del}) and (\ref{F}). The nontrivial solution of ${\bf \Delta}$ requires
the determinant of the coefficient matrix being zero. This way to solve numerically the BCS Hamiltonian is exact in the mean field level.

\subsection{Single-wall carbon nanotubes}
In order to give an analytic solution, let us consider the SWCNs first. 
We note that the density of states (DOS) of metallic tubes per atom can 
be expressed approximately near Fermi level as \cite{Saito,White} 
\begin{equation}
\rho(E_{F})=\frac{4a}{\pi^{2}td},
\label{rof}
\end{equation}
where $d$ is the diameter of the tube. Thus, the DOS near Fermi level depends on the diameter of the tubes.
The equation of the superconducting energy gap may be written as
\begin{equation}
1=V\rho(E_{F})F_{0}(\Delta)
\label{gap0}
\end{equation}
Since the Debye energy of the phonons $\hbar\omega_{D}$ is estimated to be $0.1eV$ \cite{Gonzalez}, 
we may neglect the effect of the energy band structure beyond the Fermi level. Thus, the function $F_{0}(\Delta)$ 
may be written as
\begin{equation}
F_{0}(\Delta)=\int_{0}^{\hbar\omega_{D}} 
\frac{d\epsilon}{\sqrt{\epsilon^{2}+\Delta^{2}}}
\tanh\left( \frac{\sqrt{\epsilon^{2}+\Delta^{2}}}{2k_{B}T}\right) 
\label{FD}
\end{equation}
At the critical temperature, $T=T_{c}$, the superconducting paring
equals to zero, $\Delta=0$. The integration Eq.(\ref{FD}) can be integrated  
\begin{equation}
F_{0}(0)=\ln\left(\frac{2e^{\gamma}}{\pi}\frac{\hbar\omega_{D}}{k_{B}T_{c}}\right)
\label{F0}
\end{equation}
where $\gamma \approx 0.5772$ is Euler constant. Using Eq.(\ref{rof}), (\ref{gap0}), and (\ref{F0}), 
we can obtain the superconducting transition temperature,
\begin{equation}
k_{B}T_{c}=\frac{2e^{\gamma}}{\pi}\hbar\omega_{D}\exp\left(-\frac{\pi^{2}td}{4aV}\right)
\label{Tc1}
\end{equation}
Interestingly, the critical temperature $T_{c}$ decreases exponentially with the increase of the diameter
of the tube. We also obtain the superconducting paring at zero temperature is 
$\Delta(0)=2\hbar\omega_{D}\exp\left(-\frac{\pi^{2}td}{4aV}\right)$. Thus, the superconducting paring
can be also expressed approximately as
\begin{eqnarray}
\Delta(T)\approx\left\{
\begin{array}{ll}
\Delta(0)-(2\pi k_{B}T\Delta(0))^{1/2}e^{-\frac{\Delta(0)}{k_{B}T}} & {\rm for\   } T\ll T_{c} \\
3.06k_{B}T_{c}(1-\frac{T}{T_{c}})^{1/2}  &  {\rm for \   }  T\rightarrow T_{c}
\end{array}
\right.
\end{eqnarray} 
Since there is an energy gap
at Fermi level for the semiconducting SWCNs, the superconducting state cannot occur in the semiconducting SWCNs.
However, it has been found that the layer-layer coupling of MWCNs may induce the semiconductor-metal phase 
transition.\cite{Liang,Kociak} Actually, most of MWCNs are metallic due to the interlayer coupling.\cite{Liang2,Kociak} 
Thus, the superconducting state may occur in MWCNs.  

\subsection{Double-wall carbon nanotubes}
For the double-wall carbon nanotubes (DWCN), the Cooper paring equation Eq. (\ref{Del}) 
can be reduced to \cite{George,Yuanhe} 
\begin{equation}
\left( 
\begin{array}{cc}
1-V_{1}\rho_{1}(E_{F})F_{0}(\Delta_{1}) & -V_{1,2}\rho_{2}(E_{F})F_{0}(\Delta_{2})  \\
-V_{2,1}\rho_{1}(E_{F})F_{0}(\Delta_{1}) & 1-V_{2}\rho_{2}(E_{F})F_{0}(\Delta_{2})  \\
\end{array}
\right)
\left( 
\begin{array}{c}
\Delta_{1} \\
\Delta_{2} \\
\end{array}
\right)=0,
\label{F2} 
\end{equation}
The effective electron-electron interaction may be assumed to be independent of the chirality of the tube, namely
$V_{1}=V_{2}\equiv V$ and let $V_{1,2}=V_{2,1}\equiv V_{\perp}$. 
At $T=T_{c}$, the superconducting paring equals to zero, $(\Delta_{1},\Delta_{2})=0$, 
The nontrivial solution of Eq.(\ref{F2}) requires the determinant of the coefficient matrix being zero, 
in which substituting Eq.(\ref{rof}) into Eq.(\ref{F2}), we can solve two solutions of $F_{0}(0)$ of 
Eq.(\ref{F2}). Based on the hints of the experimental results \cite{Kasumov,Takesue} 
and the similar two-band BCS theory,\cite{George} we may select the smallest solution of $F_{0}(0)$, 
\begin{equation}
F_{0}(0)=\frac{\pi^{2}t}{8(V^{2}-V_{\perp}^{2})a}[V(d_{1}+d_{2})-
\sqrt{V^{2}(d_{2}-d_{1})^{2}+4V_{\perp}^{2}d_{1}d_{2}}].
\end{equation}  
Combining Eq.(\ref{F0}), the superconducting transition temperature can be obtained 
\begin{equation}
k_{B}T_{c}=\frac{2e^{\gamma}}{\pi}\hbar\omega_{D}e^{-F_{0}(0)}.
\label{Tc2}
\end{equation}

Thus, the interlayer hopping of the Cooper pair enhances the 
superconducting paring and increases the superconducting transition temperature, 
which is consistent with the experimental observation\cite{Kasumov,Takesue}.

\subsection{Multi-wall carbon nanotubes}
Similarly, we can generalize above way of the DWCN to the MWCN case. Suppose MWCN containing $M$ layers,  
to facilitate the analytical solution of $F_{0}(0)$,
we assume that $V_{\ell}\rho_{\ell}(E_{F})= V_{\ell'}\rho_{\ell'}(E_{F})\equiv V\langle\rho(E_{F})\rangle$
and  $V_{\ell,\ell+1}\rho_{\ell+1}(E_{F})= V_{\ell+1,\ell}\rho_{\ell}(E_{F})\equiv V_{\perp}\langle\rho(E_{F})\rangle$,
where  $\langle\rho(E_{F})\rangle=\frac{4a}{\pi^{2}t\langle d\rangle}$. The $\langle d\rangle$ is the average of the 
diameter of MWCN, $\langle d\rangle=\frac{1}{M}\sum_{\ell}d_{\ell}$.
At $T=T_{c}$, the determinant of the coefficient of the Eq.(\ref{Del}) can be reduced to the Chebyschev polynomial, 
which can be solved,\cite{George}
\begin{equation}
F_{0}(0)=\frac{1}{\langle\rho(E_{F})\rangle[V+2V_{\perp}\cos(\frac{\pi}{M+1})]},
\end{equation}  
where we has also selected the smallest solution of the Chebyschev polynomial like the DWCN case.\cite{George}
The superconducting transition temperature can be expressed as 
\begin{equation}
k_{B}T_{c}=\frac{2e^{\gamma}}{\pi}\hbar\omega_{D}
\exp\left(-\frac{\pi^{2}t\langle d\rangle}{4a[V+2V_{\perp}\cos\left(\frac{\pi}{M+1}\right)]}\right)
\label{Tm}
\end{equation}

It can be also seen that the Cooper pair tunnelling between layers will enhacnce the superconducting correlation 
and increase the critical temperature. This agrees with the experimental observation.\cite{Kasumov,Takesue}
This theoretical framework provides us an physical understanding of the superconductivity mechanism of CNs
even though the result relies on the $F_{0}(0)$ solution selected based on the hints of 
the experiment result\cite{Kasumov,Takesue} and the multi-band BCS theory.\cite{George}
 
Experimentally, the transport measurement of the ropes of SWCNs indicates the superconducting state occuring
in the temperature range of $0.1\sim 1K$.\cite{Kasumov} The investigation of the magnetic 
properties for the small radius SWCN ($4\AA$) reveals that the $T_{c}$ can be estimated to be lower than $15K$.\cite{Tang}
The transport measurements of the $Au/MWCN/Au$ junctions indicate that the $T_{c}$ reaches to $6K\sim12K$.\cite{Takesue}
The theoretical study of the zigzag and armchair SWCNs by the Millan's formula and the $VASP$ software
package gives that the $T_{c}$ is very small, $10^{-9}K$ for the armchair tube $(5,5)$,
but $55\sim75K$ for the zigzag tube $(5,0)$.\cite{Iyakutti} 

As an example of this formalism we list several typical results of the superconducting transition temperature
in the table I and compare the first-principle calculation\cite{Takesue} and the experimental results,\cite{Kasumov,Tang,Takesue}
where we use the parameters $t=2.5eV$, $V=1.2eV$, and $V_{\perp}=0.8eV$, $\hbar\omega_{D}=0.1eV$.\cite{Gonzalez} 
For SWCN the first-principle calculation predicts the (5,0) tube being metallic due to  the $\pi^{*}-\sigma$ coupling.\cite{Iyakutti} 
However, in the $\pi$ electronic tight-binding approximation the (5,0) tube is semiconducting.\cite{Saito}  Thus,
we chose a small diameter tube (4,1) ($\sim4\AA$) for comparision with the results of the experimental observation
\cite{Tang} and the first-principle calculation.\cite{Iyakutti} 
The result we obtain is $T_{c}=0.729K$, which agrees qualitatively with the experimental result.\cite{Kasumov,Takesue,Tang} 
For SWCN (5,5), we estimate the $T_c=9.23\times10^{-4}K$, which is the same order to another theoretical results,
$T_c=1.5\times10^{-4}K$.\cite{Yuanhe} In their theory  
the electron-phonon interaction is taken into account in detail within the BCS framework. 
For MWCN the Cooper pair tunnelling between layers also enhances the Cooper pair correlation 
and increase the superconducting transition temperature. 
This is also consistent with the experimental observation.\cite{Kasumov,Takesue}

\begin{table}
\caption{\label{tab:table1}The superconducting transition temperature of several CNs.} 
\begin{tabular}{lllll}
\hline\hline
CNs &  (n,m)@(n,m)... & $T_{c}(K)^{\footnotemark[1]}$ & $T_{c}(K)^{\footnotemark[2]}$ & $T_{c}(K)^{\footnotemark[3]}$ \\
\hline
SWCN & $(4,1)$  & 0.729 & 57 & $<15$ \\
SWCN & (5,5) & $9.23\times 10^{-4}$ &$3\times 10^{-9}$ & $0.1\sim 1$(SWCN rope)\\
SWCN & (10,10) & $6.46\times 10^{-10} $ &$2.6\times 10^{-30}$ &\\
DWCN & (5,0)@(10,0) & 2.24 & &\\
DWCN & (5,5)@(8,8) & 0.043 & &\\
TWCN & (5,0)@(10,0)@(15,0) & 0.29 & & $6\sim12 $(MWCN junction)\\
TWCN & (5,5)@(8,8)@(12,12) & 0.0069 & &\\
\hline\hline
\end{tabular}
{$^{1}$our results; $^{2}$the first-priniciple calculation\cite{Iyakutti}; $^{3}$ experimental results.\cite{Kasumov,Takesue,Tang}}
\end{table}

\section{Discussion and conclusion}
Physically, the intratube electronic hopping constant $t$, the effective electron-electron interaction $V_{\ell}$
and the intertube electron-electron interaction $V_{\ell,\ell'}$ could depend slightly on the chirality and the layer number 
of the tube.\cite{Gonzalez,Yuanhe} Thus, We cannot obtain exactly an analytical solution of $T_{c}$. 
Nevertheless, above formalism Eqs.(\ref{H})$\sim$ (\ref{Del}) still provides a guideline for the numerical study. 
From the theoretical point of views, we give the BCS formalism on MWCN and the formula in Eq. (\ref{Tm}) tell us that
the Cooper pair tunnelling between layers enhances the superconducting correlation and $T_{c}$, which is cconsistent
with the experimental results\cite{Kasumov,Takesue} even though we make some approximations in Eq.(\ref{Tm}) for obtaining an analytic formula.   
Actually, for MWCN the effect of the layer number on $T_{c}$ competes with the effect of the diameter of MWCN, which can be seen from Eq.(\ref{Tm}).
However, the effect of diameter dominates $T_{c}$ for the MWCNs that have the layer number more than 4. Hence, the small-diameter 
double-wall or triple-wall carbon nanotubes could have higher $T_{c}$. 

On the other hand, the study of the thermodynamic variables starts from the partition function of the system, $Z=Tr(e^{-\beta H})$,
where $\beta$ is the inverse temperature. The free energy may be expressed in terms of $F=-\beta\ln Z$. Using Eq.(\ref{Hk}),
we can obtain
\begin{equation}
F=-\frac{2}{\beta}\sum_{k,\ell}\left[\ln(1+e^{-\beta \xi_{\ell}(k)})
-\epsilon_{\ell}(k)+\xi_{\ell}(k)-\frac{\Delta^{2}_{\ell}}{2\xi_{\ell}(k)}\right].
\label{FE}
\end{equation}
The other thermodynamic variables can be obtained by the thermodynamic relationships between the free energy
and the thermodynamic variables, such as the electronic entropy,
$S=-\frac{\partial F}{\partial T}$, and the specific heat $C=T\frac{\partial S}{\partial T}$ .\cite{BCS}

In summary, we have developed a BCS formalism of superconductivity for MWCNs, which give that 
the superconducting transition temperature $T_{c}$ of SWCNs decreases exponentially the tube diameter. 
The $T_{c}$ is very small (less than 1K for the diameter less than 1nm).
The Cooper pair tunnelling between layers enhances the superconducting correlation and 
increases the supersonconducting transition temperature. The interlayer coupling increases the number 
of the transverse channels, which suppresses 1D characteristic of CNs due to screening the repulsive 
electron-electron interaction. Hence, the superconducting long-range order emerges favorably in 
the small-diameter double-wall or triple-wall carbon nanotubes . 

{\bf Acknowledgments}
The author gratefully acknowledge the financial support of the project from the National Natural Science Foundation of China 
(Grants No. 10774194; 90306016; 50572123), National Basic Research Program of China (973 Program: 2007CB935501),
Advanced Academic Research Center of Sun Yat-Sen university (06P4-3), and SRF for ROCS SEM.


\end{document}